\documentclass[journal]{IEEEtran}
\usepackage{amsmath,graphicx,amssymb}
\usepackage{bm}
\usepackage{csvsimple}
\usepackage{pgfplots}
\usepackage{tikz}
\usetikzlibrary{positioning}
\usepackage{tumcolor}
\usepackage{amsmath}
\usepackage{amsthm,amssymb}

\usepackage[noadjust]{cite}

\usepackage[ruled,linesnumbered]{algorithm2e}

\DeclareMathOperator{\Hermitian}{H}
\newcommand{\He}{{\Hermitian}}
\DeclareMathOperator{\Transpose}{T}
\newcommand{\Tr}{{\Transpose}}

\DeclareMathOperator{\trace}{tr}
\DeclareMathOperator{\diag}{diag}
\DeclareMathOperator{\bdiag}{blockdiag}

\DeclareMathOperator{\Exp}{E}
\DeclareMathOperator{\modulo}{mod}
\DeclareMathOperator*{\argmax}{argmax}

\DeclareMathOperator*{\st}{s.t.\!:}

\definecolor{MyPurple}{rgb}{0.4,0,0.8}
\definecolor{MyGreen}{rgb}{0,0.5,0}
\definecolor{MyBrown}{rgb}{0.5,0.3,0.1}
\definecolor{MyCyan}{rgb}{0.2,1,1}
\definecolor{MyPink}{rgb}{1,0.2,1}
\usepackage{hyperref}
\newcommand\copyrighttext{%
	\footnotesize \textcopyright 2021 IEEE. Personal use of this material is permitted.
	Permission from IEEE must be obtained for all other uses, in any current or future
	media, including reprinting/republishing this material for advertising or promotional
	purposes, creating new collective works, for resale or redistribution to servers or
	lists, or reuse of any copyrighted component of this work in other works.
	DOI: \href{https://ieeexplore.ieee.org/document/9432742}{10.1109/TSP.2021.3081047}}
\newcommand\copyrightnotice{%
	\begin{tikzpicture}[remember picture,overlay]
	\node[anchor=south,yshift=10pt] at (current page.south) {\fbox{\parbox{\dimexpr\textwidth-\fboxsep-\fboxrule\relax}{\copyrighttext}}};
	\end{tikzpicture}%
}

\title{A Machine Learning Approach to DoA Estimation and Model Order Selection for Antenna Arrays with Subarray Sampling}
\author{Andreas Barthelme, Wolfgang Utschick\\Professur f\"ur Methoden der Signalverarbeitung, Technical University Munich, 80290 Munich, Germany\\Email: \{a.barthelme, utschick\}@tum.de}

\begin{document}
%
\maketitle
\copyrightnotice
\begin{abstract}
In this paper, we study the problem of direction of arrival estimation and model order selection for systems employing subarray sampling. Thereby, we focus on scenarios, where the number of resolvable active sources is not smaller than the number of simultaneously sampled antenna elements, i.e., we operate above the conventional limit for most estimators. For this purpose, we propose new schemes based on neural networks and estimators that combine neural networks with gradient steps on the likelihood function. These methods are able to outperform existing estimators in terms of mean squared error and model selection accuracy, especially in the low snapshot domain, at a drastically lower computational complexity.
\end{abstract}
\section{Introduction}
\label{sec:intro}
In the last decade, data-driven approaches have become increasingly popular in the signal processing community. Driven by astonishing results from image and speech processing, deep neural networks have become a tool that finds its way to many different signal processing applications. Wherever there are model imperfections or the existing solutions are very complex to compute, data-based approaches may improve the performance considerably. 

Traditionally, direction of arrival (DoA) estimation is a field where appropriate stochastic models and potent algorithms are available. However, there are still some areas where existing solutions lead to a rather limited performance. Subarray sampling is one of these applications for which classical methods do not provide fully satisfying results. The idea behind subarray sampling is to reduce costs by sequentially sampling subarrays instead of sampling the whole antenna array simultaneously, which means that fewer radio frequency (RF) chains than antennas are needed. Specifically, in the domain where the number of sources is not smaller than the number of sampled antenna elements per time step \cite{Sheinvald1999,Fishler2000}, the performance of existing DoA estimation algorithms is---as we will show---only acceptable for a prohibitively high number of snapshots. Therefore, we investigate the suitability of machine learning--based approaches for systems with subarray sampling. In particular, we discuss neural networks (NNs) for the tasks of DoA estimation and model order selection. 

Subarray sampling can be seen as a special form of time-varying arrays. Computable methods for DoA estimation for time-varying arrays fall into two categories depending on the ratio of transmitting sources to simultaneously sampled antenna elements. For fewer sources than sampled antennas per time step, previous work goes back to \cite{Zeira1995}, where the single source case is studied. In \cite{Friedlander1996}, the same authors extend their analysis of time-varying arrays to multiple sources and propose eigenstructure methods based on array interpolation and focusing matrices. A more direct approach to employing MVDR and MUSIC for these systems is studied in \cite{Rieken2004,Worms2000}. The more demanding scenario, where the number of sources is equal or greater than the number of simultaneously sampled antennas, is discussed in \cite{Sheinvald1999}. There, the authors propose to use a cost function that matches the subarray covariance matrices to the observed sample covariance matrices in a general least squares (GLS) sense. In \cite{Fishler2000}, the covariance matrix of the full antenna array is estimated by a special subarray sampling scheme. Afterwards, the DoAs are estimated from the reconstructed, full covariance matrix with MUSIC \cite{Schmidt1986}. Lastly, the recent work in \cite{Suleiman2018} on non-coherent processing of partly calibrated arrays yields an estimator that utilizes a sparse signal representation of the system model and is applicable to subarray sampling.

Utilizing data-based machine learning techniques for DoA estimation goes back to the '80s to Rastogi et al.~\cite{Rastogi1987}. A review of the work from the last century in this field can be found in \cite{Du2002}. More recently, with the increase in computing resources, the methods have shifted towards larger fully connected and convolutional multilayer NNs. The proposed NN approaches can be assigned to three different groups. One group poses the DoA estimation problem as a classification problem by a discretization of the angular domain in several sectors (e.g., \cite{Liu2018,Chakrabarty2019,Ozanich2020,Yao2020}). The DoA estimation problem then reduces to determining if a source is present in a specific angular sector. For the next group, the idea is to estimate a discretized spectrum (MUSIC \cite{Elbir2020} or transmit power \cite{Wu2019}) as a proxy by means of a regression network, and derive the DoA estimates by determining the maxima of the respective spectrum. In \cite{Izacard2019}, such a NN-based estimation of a pseudo-spectrum has been shown to perform better than directly estimating the frequencies with a NN in a multisinusoidal signal. A disadvantage of the aforementioned methods is that a minimal angular spread between two sources has to be assumed, such that each grid point or sector is only associated with one possible source. The third group, which our proposed method belongs to, does not suffer from this. There, the NN should directly produce the DoA estimates at its output. These models are trained directly on the cost function of interest, e.g., mean squared error (MSE). In \cite{Guo2020}, the authors propose to use a signal-to-noise-ratio (SNR) classification network to choose between two different DoA regression networks with the goal to resolve two narrowly spaced sources. A more general approach is presented in \cite{Bialer2019}. There, a NN is discussed that is able to simultaneously estimate the number of sources and their DoAs. The authors show that this network is able to achieve the same performance as a maximum likelihood (ML) estimator in a scenario with two sources and a single snapshot.

For model order selection, NNs have been first proposed in \cite{CostaHirschauer1995}. Recently, the model order selection problem has been revisited for state of the art fully connected, feedforward network architectures with different input data formats \cite{Bialer2019,Yang2020,Barthelme2020}. In the context of subarray sampling, previous work on model order selection is limited to \cite{Haykin1995}, where the applicability of information criteria to the time-varying preprocessing case is discussed.

In this paper, we discuss several DoA estimation methods for systems with subarray sampling in Section \ref{sec:doachap}. There, our contributions lie in the proposal of a NN-based DoA estimator and the modification of the GLS estimator for a small number of snapshots. Moreover, we provide a thorough comparison of the newly proposed and existing schemes for the critical case of as many sources as simultaneously sampled antennas by means of Monte Carlo simulations. These simulations show that the proposed NN-based estimation scheme is able to outperform the state-of-the-art estimators in terms of estimation accuracy and computational complexity. In Section \ref{sec:modelorder}, the model order selection problem for subarray sampling is discussed. We present a new estimation scheme for the model order based on a NN and are---to the best of our knowledge---the first to provide simulation results for the achievable selection accuracy for these systems. Again, the proposed NN-based approach is able to provide a significantly better performance compared to existing methods based on information criteria, as it provides a higher selection accuracy at a fraction of the computational cost.

\section{System Model}
\label{sec:sysmodel}
Let us consider antenna arrays consisting of $M$ antennas. Throughout this work, we investigate systems which only use $W<M$ RF chains and a switching network to sample the received signals, i.e., only a subset of the antenna elements is sampled simultaneously. In the following, we assume that at any given time each RF chain is connected to exactly one antenna element and that there are $K$ different states of the switching network, i.e., we have $K$ different subarrays consisting of $W$ antennas. For each subarray, we collect $N$ snapshots for a joint processing. Then, the $n$-th sample of the received signal for the $k$-th subarray can be written as
\begin{equation}
\bm{y}^{(k)}(n) = \bm{G}^{(k)}\left(\bm{A}(\bm{\theta})\bm{s}^{(k)}(n)+\bm{\eta}^{(k)}(n)\right),\label{eq:sysmodel}
\end{equation} 
where the steering matrix $\bm{A}(\bm{\theta})\in\mathbb{C}^{M\times L}$ captures the response of the whole array on the DoAs $\bm{\theta}$ of $L$ far-field sources, $\bm{s}^{(k)}$ denotes the narrow-band transmit signals, and $\bm{\eta}^{(k)}\sim\mathcal{CN}(\mathbf{0},\sigma_\eta^2\mathbf{I}_M)$ is some additive white Gaussian noise. The matrix $\bm{G}^{(k)}\in\{0,1\}^{W\times M}$ represents the connections between the RF chains and the antenna elements that form the $k$-th subarray.

\section{DoA Estimation}
\label{sec:doachap}
In this section, we briefly discuss existing DoA estimation approaches that are derived from the underlying stochastic model, before we present a new data-driven NN approach. Afterwards, we compare these methods by means of Monte Carlo simulations.
\subsection{Model-Based DoA Estimation}
\label{sec:doa}
Traditionally, DoA estimation methods have been derived from the underlying stochastic model. In the scope of DoA estimation, two different stochastic models are commonly associated with the received signals $\bm{y}$ given in (\ref{eq:sysmodel}). These models differ in the assumed distribution of $\bm{s}$ \cite{Krim1996}. On the one hand, we may treat $\bm{s}$ as an unknown parameter of the stochastic model. On the other hand, we may assume that $\bm{s}$ itself follows some probability distribution, which leads to a stochastic model for $\bm{y}$ that no longer depends on the individual realizations of $\bm{s}$, but on the parameters of its distribution. 

\subsubsection{Maximum Likelihood Estimator}
As the most prominent model-based estimation method, we start by discussing the ML estimator. The ML estimator finds its estimates by maximizing the probability density function at the observed received signals $\bm{y}$ with respect to the model parameters. Deriving the ML estimator under the aforementioned stochastic models, we obtain the deterministic ML (DML) for the former model and the stochastic ML (SML) for the latter model. In the case of $L<W$, i.e., fewer sources than RF chains, the computation of the DML estimates is straightforward. As it is well summarized in \cite{Zeira1995a}, we can find closed form estimates of the signal and noise parameters for fixed angles $\bm{\theta}$. Plugging these estimates into the likelihood function gives a non-convex function in $\bm{\theta}$. To find the global maximum of this concentrated likelihood, a multi-dimensional grid search over $\bm{\theta}$ followed by any kind of gradient ascent technique can be employed.

More challenging is the case with an equal to or greater number of sources than RF chains $L\geq W$, which we will focus on throughout this section. For the DML case, we have 
\begin{equation}
	\bar{\bm{y}}(n)=\begin{bmatrix}
	\bm{y}^{(1)}(n)\\\vdots\\\bm{y}^{(K)}(n)
	\end{bmatrix}\sim\mathcal{CN}\left(\bar{\bm{A}}\bar{\bm{s}}(n),\mathbf{I}_{KW}\right),
\end{equation}
with 
\begin{align}
\bar{\bm{A}}(\bm{\theta})&=\bdiag\left\{\bm{G}^{(1)}\bm{A}(\bm{\theta}),\dots,\bm{G}^{(K)}\bm{A}(\bm{\theta})\right\},\\
\bar{\bm{s}}(n)&=\left[\bm{s}^{(1),\He}(n),\dots,\bm{s}^{(K),\He}(n)\right]^\He.
\end{align}
Since in general $\bar{\bm{A}}(\bm{\theta})$ does not have full column rank for $L\geq W$, there is a manifold of solutions for $\bm{\theta}$ and $\bar{\bm{s}}(n)$ that give the same distribution for $\bar{\bm{y}}(n)$, i.e., $\bm{\theta}$ cannot be uniquely estimated with the DML model \cite{Hochwald1996}. 

In contrast, the optimization problem corresponding to the SML estimator, where $\bm{s}(t)\sim\mathcal{CN}(\mathbf{0},\bm{R}_{\bm{s}})$, cannot be reduced to an optimization which only depends on $\bm{\theta}$ for $L>1$ \cite{Friedlander1996}. Instead, the SML likelihood estimates are the solution to
\begin{equation}
	\max_{\bm{\theta},\bm{R}_{\bm{s}}\succ\mathbf{0},\sigma_\eta^2\geq\mathbf{0}}-\sum\limits_{k=1}^K\left[\ln\left(\det(\bm{R}^{(k)}_{\bm{y}})\right)+\trace(\bm{R}^{(k),-1}_{\bm{y}}\hat{\bm{R}}^{(k)}_{\bm{y}})\right],\label{eq:SMLobjective}
\end{equation}
with the covariance matrices
\begin{equation}
	\bm{R}^{(k)}_{\bm{y}}=\bm{G}^{(k)}\bm{A}(\bm{\theta})\bm{R}_{\bm{s}}\bm{A}^\He(\bm{\theta})\bm{G}^{(k),\Tr}+\sigma_\eta^2\mathbf{I}_W,
\end{equation}
and the sample covariance matrices
\begin{equation}
	\hat{\bm{R}}^{(k)}_{\bm{y}}=\frac{1}{N}\sum\limits_{n=1}^N\bm{y}^{(k)}(n)\bm{y}^{(k),\He}(n).
\end{equation}
This optimization problem has almost surely a unique maximizer if $\bm{R}_{\bm{s}}$ is diagonal and $L\leq\lfloor\frac{\rho}{2}\rfloor$ \cite{Suleiman2018}, where $\rho$ is the Kruskal rank of the co-array manifold
\begin{equation}
	\breve{\bm{V}}(\bm{\theta})=\begin{bmatrix}\left(\bm{G}^{(1)}\bm{A}^{*}(\bm{\theta})\right)\circ\left(\bm{G}^{(1)}\bm{A}(\bm{\theta})\right)\\\vdots\\\left(\bm{G}^{(K)}\bm{A}^{*}(\bm{\theta})\right)\circ\left(\bm{G}^{(K)}\bm{A}(\bm{\theta})\right)\end{bmatrix},
\end{equation}
that uses the Khatri-Rao product, denoted by $\circ$. For the correlated source case with a dense covariance matrix $\bm{R}_{\bm{s}}$, the extension of the identifiability proof in \cite{Suleiman2018} is non-trivial, and, as of yet, remains an open problem.

Unfortunately, for the optimization of (\ref{eq:SMLobjective}), we cannot find closed form solutions for $\bm{R}_{\bm{s}}$ and $\sigma_\eta^2$ for fixed $\bm{\theta}$. Therefore, the optimization of this non-convex function is over $L^2+L+1$ variables, which is computationally very expensive for any $L\geq 2$. To overcome this problem, different methods that replace the likelihood objective with a covariance-matching criterion have been proposed to estimate $\bm{\theta}$ specifically for $L\geq W$ \cite{Sheinvald1999,Fishler2000,Suleiman2018}. As the method introduced in \cite{Fishler2000} only works for a special sampling scheme, in which every lag in the covariance matrix needs to be sampled by at least one of the subarrays. Instead, we focus on the GLS \cite{Sheinvald1999} and sparse signal repesentation (SSR) \cite{Suleiman2018} methods that do not suffer from this restriction.

\subsubsection{GLS Estimator}
\label{sec:GLS}
The GLS estimator has been shown to be an asymptotically consistent and efficient estimator. The idea is to obtain the estimates by a covariance fitting criterion. The GLS approach solves the following optimization problem
\begin{equation}
	\min_{\bm{\theta},\bm{R}_{\bm{s}}\succeq\mathbf{0},\sigma_\eta^2\geq 0}\sum\limits_{k=1}^K\|\bm{T}^{(k)}\left(\hat{\bm{R}}^{(k)}_{\bm{y}}-\bm{R}^{(k)}_{\bm{y}}\right)\bm{T}^{(k),\He}\|_{\text{F}}^2,\label{eq:GLSobjective}
\end{equation}
where $\bm{T}^{(k)}$ is a whitening filter, for which the choice $\bm{T}^{(k)}=\hat{\bm{R}}_{\bm{y}}^{(k),-1/2}$ can be motivated by asymptotic considerations \cite{Sheinvald1999}. In the optimization above, the signal and noise estimates can be computed for fixed $\bm{\theta}$. Hence, the non-convex optimization problem results in a search for the optimal $\bm{\theta}$, which can again be solved by a grid-search approach.

Note that in contrast to the original paper \cite{Sheinvald1999}, we include the positive-semidefiniteness constraints on $\bm{R}_{\bm{s}}$ and $\sigma_\eta^2$ in the optimization problem (\ref{eq:GLSobjective}). Otherwise, we obtain some infeasible results for the estimates of $\bm{R}_{\bm{s}}$ and $\sigma_\eta^2$ when working in the low snapshot domain. This, in turn, means that the effort for determining the noise and signal estimates for fixed $\bm{\theta}$ is not a simple least squares problem, but requires the solution of a semidefinite program in the general case and non-negative least squares problem (quadratic program) in the case of uncorrelated sources.

\subsubsection{SSR Estimator}
The SSR estimator has been derived in \cite{Suleiman2018} from the SPICE estimator \cite{Stoica2011} for DoA estimation in partly calibrated arrays. Due to its non-coherent processing, i.e., phase offsets between the subarrays are not estimated, it is directly applicable to the subsampling system model at hand. Similar to the GLS approach, the SSR method is based on a covariance-matching cost function given by
\begin{equation}
\sum\limits_{k=1}^K\|\bm{R}^{(k),-1/2}_{\bm{y}}\left(\hat{\bm{R}}^{(k)}_{\bm{y}}-\bm{R}^{(k)}_{\bm{y}}\right)\hat{\bm{R}}^{(k),-1/2}_{\bm{y}}\|_{\text{F}}^2.\label{eq:SSRobjective}
\end{equation}
Using a sparse representation of the covariance matrices $\bm{R}^{(k)}_{\bm{y}}$ for uncorrelated signals, a second-order cone program (SOCP) can be derived from (\ref{eq:SSRobjective}). Although the derivation of the SSR estimator is based on the assumption of uncorrelated signals, the authors argue that due to the robustness of sparse signal models the method can also be used for the correlated case. The resulting SOCP can be either solved by a general purpose solver or, as has been proposed for SPICE in \cite{Stoica2011}, a cost-effective alternating optimization method can be used (for details see Appendix \ref{sec:altopt}).

\subsection{Purely Data-Based DoA Estimation}
In contrast to the previously discussed model-based methods, in this subsection, we discuss a purely data-based DoA estimation approach. In particular, we present an end-to-end, feedforward NN that is trained on artificial training data sampled from the SML signal model. In the following, we refer to this NN as \emph{MCENet}, named after the employed objective function. Our main focus lies on the uncorrelated case, i.e., $\bm{R}_{\bm{s}}$ is diagonal, because of multiple reasons. This assumption provides sufficient identifiability conditions, is at the core of the SSR estimator, and reduces the complexity of the GLS estimator. Note that an extension of the proposed scheme to the correlated case simply means sampling data from the respective stochastic model.

\subsubsection{Data and preprocessing}
\label{sec:data}
For our training set, we sample data from the system model in (\ref{eq:sysmodel}). Hereby, the entries of the DoA vector $\bm{\theta}$ are drawn from a uniform distribution over the complete field of view, i.e., $\theta\sim\mathcal{U}(0,U)$. For now, let us assume that the entries in $\bm{\theta}$ are sorted in ascending order, which will be discussed later. The noise and transmit signal realizations are drawn from uncorrelated Gaussian distributions according to the SML model. Thereby, we fix the power of the strongest source to $\sigma_{s,\text{max}}^2=1$, and for each realization, we draw the power of each weaker source in decibel from a uniform distribution between $0\,\text{dB}$ and $\sigma_{s,\text{min}}^2$. The noise power also follows a uniform distribution between $\sigma_{\eta,\text{min}}^2$ and $\sigma_{\eta,\text{max}}^2$. Each data sample consists of $KN$ i.i.d.~received signal realizations $\bm{y}^{(k)}(n),n=1,\dots,N, k=1,\dots,K$. Since we use artificial data, we can feed new, previously unseen realizations to the NN in each step of the gradient descent of the learning algorithm. This makes the training inherently robust towards overfitting. Additionally, we know the true DoAs for each realization, therefore, we can use them as the label for each data sample in a supervised learning approach.

We pass sample covariance matrix information to the input layer of the NN, which has been shown to be a good preprocessing step for the complex received signals in the DoA context \cite{Costa1999,Niu2017,Barthelme2020}. To this end, we form the $K$ subarray sample covariance matrices $\hat{\bm{R}}_{\bm{y}}^{(k)}$, $k=1,\dots,K$, and stack their real parameters, i.e., their diagonal elements and the real and imaginary parts of their upper triangle, in one large vector per data sample. Note that the input size of the NN is $KW^2$, and thus does not depend on the number of snapshots $N$.

\subsubsection{Architecture and cost function}
\label{sec:architecture} 
Due to its simplicity, we use a fully connected, feedforward NN with $N_{\text{h}}$ hidden layers, each consisting of $N_{\text{u}}$ neurons. For the non-linear activation function of the hidden layers, we employ the rectified linear unit (ReLU). The output layer produces $L$ outputs, which are the estimates of the DoA $\hat{\bm{\theta}}$. 

The most common cost function for parameter estimation is the MSE. However, for DoA estimation the $2\pi$-periodicity of the angles has to be taken into account. To that end, the mean squared periodic error (MSPE), given by
\begin{equation}
	\text{MSPE}(\theta,\hat{\theta})=\Exp_\theta\left[\left|\modulo_{[-\pi,\pi)}\left(\theta-\hat{\theta}\right)\right|^2\right],
\end{equation} 
has been proposed \cite{Reggiannini1997}. An alternative, which is differentiable at every point and is equivalent to the MSE in the small error region, is the mean cyclic error (MCE) \cite{Routtenberg2013}. The MCE can be calculated according to
\begin{equation}
	\text{MCE}(\theta,\hat{\theta})=\Exp_\theta\left[2\left(1-\cos\left(\theta-\hat{\theta}\right)\right)\right].
\end{equation}
Although the non-differentiability of the MSPE is only at one point, and hence, can be simply replaced by its left derivative without any adverse impact on the learning procedure, we use the MCE with its continuous derivative for the cost function of the NN.

For $L>1$, the order of the elements in $\hat{\bm{\theta}}$ should be irrelevant for the value of the cost function. Therefore, the minimum of the sum of the element-wise errors between the true DoA and all permutations of $\hat{\bm{\theta}}$ is used for the cost function $f(\bm{\theta},\hat{\bm{\theta}})$, i.e.,
\begin{equation}
	f(\bm{\theta},\hat{\bm{\theta}})=\min_{\bm{\Pi}} \sum_{\ell=1}^L f\left(\theta_\ell,\bm{\pi}_\ell^\Tr\hat{\bm{\theta}}\right),
	\label{eq:permutatedcost}
\end{equation}
where $\bm{\Pi}=[\bm{\pi}_1,\dots,\bm{\pi}_L]^\Tr$ is a permutation matrix. However, in our studies we observed that if the network is fed with sorted labels, it converges to a point where the optimal permutation matrix $\bm{\Pi}$ is constant. This means that the minimizer of (\ref{eq:permutatedcost}) for every input sample is the same, i.e., the output of the network follows a fixed order. Further studies showed that we can even omit the minimization over all permutations and simply use the sum of the element-wise errors. The network will then converge to a point, where it produces the outputs in the correct order that minimizes the sum MCE.

\subsection{Hybrid DoA Estimation}
By hybrid DoA estimation we understand the combination of two different estimation approaches in a two-stage process. In our case, this combination uses one of the model-based methods GLS and SSR or the purely data-based NN method as an initialization step and a consecutive gradient ascent method on the SML likelihood.
The $t$-th iteration of the SML gradient ascent is given by
\begin{equation}
	\bm{c}^{[t+1]} = \mathcal{P}_{\bm{c}}\left(\bm{c}^{[t]}+\alpha^{[t]}\frac{\partial \mathcal{L}(\bm{c})}{\partial \bm{c}}\Big|_{\bm{c}=\bm{c}^{[t]}}\right),
\end{equation}
where $\bm{c}^{[t]}$ gathers the values of the parameters $\bm{\theta},\bm{R}_{\bm{s}},$ and $\sigma_\eta^2$ in the $t$-th step, $\mathcal{L}(\bm{c})$ is the log-likelihood function from (\ref{eq:SMLobjective}), $\alpha^{[t]}$ is some adaptable step size, and $\mathcal{P}_{\bm{c}}(\bullet)$ indicates a projection onto the feasible set of the parameters $\bm{c}$. Thereby, the gradient of the log-likelihood $\mathcal{L}(\bm{c})$ reads according to the chain rule as
\begin{equation}
\frac{\partial \mathcal{L}(\bm{c})}{\partial c_i}=\sum_{k=1}^K\trace\left(\left(\frac{\partial \mathcal{L}}{\partial \bm{R}_{\bm{y}}^{(k)}}\right)^\He\frac{\partial\bm{R}_{\bm{y}}^{(k)}}{\partial c_i}\right),
\end{equation}
with
\begin{equation}
\frac{\partial \mathcal{L}(\bm{c})}{\partial \bm{R}_{\bm{y}}^{(k)}} = -N \left(\bm{R}_{\bm{y}}^{(k),-1}-\bm{R}_{\bm{y}}^{(k),-1}\hat{\bm{R}}_{\bm{y}}^{(k)}\bm{R}_{\bm{y}}^{(k),-1}\right),
\end{equation}
and the derivatives of the covariance matrix with respect to the individual entries of the parameter vector $\bm{c}$, which are obtained by
\begin{equation}
\begin{split}
\frac{\partial \bm{R}_{\bm{y}}^{(k)}}{\partial \theta_\ell} = &\,\bm{G}^{(k)}\bm{A}(\bm{\theta})\bm{R}_{\bm{s}}\frac{\partial}{\partial \theta_\ell}\bm{A}^\He(\bm{\theta})\bm{G}^{(k),\Tr}\\
&+\bm{G}^{(k)}\frac{\partial}{\partial \theta_\ell}\bm{A}(\bm{\theta})\bm{R}_{\bm{s}}\bm{A}^\He(\bm{\theta})\bm{G}^{(k),\Tr},
\end{split}	
\end{equation}
\begin{align}
\frac{\partial \bm{R}_{\bm{y}}^{(k)}}{\partial [\bm{R}_{\bm{s}}]_{i,j}} &= \bm{G}^{(k)}\bm{a}(\theta_i)\bm{a}^\He(\theta_j)\bm{G}^{(k),\Tr},\\
\frac{\partial \bm{R}_{\bm{y}}^{(k)}}{\partial \sigma_\eta^2} &= \mathbf{I}_W.	
\end{align}

For the model-based approaches, the consecutive gradient steps alleviate the grid mismatch problem that is inherent to any grid-based approach \cite{Yang2018}. The NN method, as posed above, does not suffer from this grid mismatch problem due to its formulation as a regression problem. However, by utilizing a purely data-based method in a scenario, where an appropriate stochastic model exists, we ignore a significant amount of information about the problem at hand. Hence, we propose the combination of NN based initialization and model aware gradient steps on the SML likelihood function to improve the DoA estimate. 

To combine the NN initialization with the SML gradient steps, an additional intermediate step is necessary. The NN does not directly yield estimates for the noise variance and signal covariance matrix, which are needed to form the initial parameter estimate $\bm{c}^{[0]}$ for the consecutive gradient approach. To that end, we propose to use the GLS estimates of these nuisance parameters $\bm{R}_{\bm{s}}$ and $\sigma_\eta^2$ for the fixed angular estimates $\hat{\bm{\theta}}$, which requires the solution of a convex optimization problem as discussed in Section \ref{sec:GLS}.

For the gradient approach on the SML likelihood, we observed that the gradient of the log-likelihood function is often dominated by the derivative w.r.t.~the noise variance $\sigma_\eta^2$ at high SNR and for $L<W$.\footnote{To explain this observation, we consider the sensitivity of the partial derivatives of the log-likelihood function in the form of its second derivatives. The expected value of the respective Hessian matrix forms the well known Fisher information matrix. For high SNR and $L<W$, the block of the Fisher information matrix that corresponds to the noise variance scales with $O(\sigma_\eta^{-4})$ in contrast to the block that corresponds to the parameters of interest $\bm{\theta}$ which grows with $O(\sigma_\eta^{-2})$ \cite[Eq. (21)]{Sheinvald1997} (Note that there is a typo in \cite[Eq. (21)]{Sheinvald1997}. The lower left entry in the matrix given for the first case should be $O(\sigma_\eta^{-4})$, as for the singular case below). } This, in turn, can lead to a slow progress in the parameters of interest, viz., the DoA estimates, during a simple gradient ascent approach. Instead, a block coordinate ascent method can be applied that alternates between updating the DoA estimates, the estimate of the signal covariance $\bm{R}_{\bm{s}}$, and an estimate of $\sigma_\eta^2$. For each these updates of the different subsets of parameters, we perform a single gradient step in the direction of the respective partial derivatives with its individual step size. This led to a much faster convergence in our simulations. The resulting structure of the hybrid estimators including the block coordinate ascent is summarized in Algorithm \ref{alg:hybrid}.

\begin{algorithm}
	\SetKwRepeat{Do}{do}{while}
	\DontPrintSemicolon
	\tcc{Intial Estimate}
	Obtain initial estimate $\hat{\bm{\theta}}^{[0]}$ by using the SSR, GLS, MCENet estimator\;
	\eIf{\textnormal{MCENet has been used}}{Obtain estimates $\bm{R}_{\bm{s}}^{[0]}$ and $\sigma_\eta^{2,[0]}$ from solving (\ref{eq:GLSobjective}) for $\bm{\theta}=\hat{\bm{\theta}}^{[0]}$\;}
	{Initial estimates $\bm{R}_{\bm{s}}^{[0]}$ and $\sigma_\eta^{2,[0]}$ have already been found in step 1\;}
	\tcc{Block Gradient Descent}
	Compute inital log-likelihood value $\mathcal{L}(\bm{c}^{[0]})$\;
	$t=0$\;
	\Do{$\mathcal{L}(\bm{c}^{[t]})>\mathcal{L}(\bm{c}^{[t-1]})$}
	{	
		$\bm{\theta}^{[t+1]} = \bm{\theta}^{[t]}+\alpha_\theta^{[t]}\frac{\partial \mathcal{L}(\bm{c})}{\partial \bm{\theta}}\Big|_{\bm{\theta}=\bm{\theta}^{[t]},\bm{R}_{\bm{s}}=\bm{R}_{\bm{s}}^{[t]},\sigma_\eta^2=\sigma_\eta^{2,[t]}}$\;
		$\bm{R}_{\bm{s}}^{[t+1]} = \bm{R}_{\bm{s}}^{[t]}+\alpha_{\bm{s}}^{[t]}\frac{\partial \mathcal{L}(\bm{c})}{\partial \bm{R}_{\bm{s}}}\Big|_{\bm{\theta}=\bm{\theta}^{[t+1]},\bm{R}_{\bm{s}}=\bm{R}_{\bm{s}}^{[t]},\sigma_\eta^2=\sigma_\eta^{2,[t]}}$\;
		$\sigma_\eta^{2,[t+1]} = \sigma_\eta^{2,[t]}+\alpha_\sigma^{[t]}\frac{\partial \mathcal{L}(\bm{c})}{\partial \sigma_\eta^{2}}\Big|_{\bm{\theta}=\bm{\theta}^{[t+1]},\bm{R}_{\bm{s}}=\bm{R}_{\bm{s}}^{[t+1]},\sigma_\eta^2=\sigma_\eta^{2,[t]}}$\;
		$t=t+1$\;
		Compute log-likelihood value $\mathcal{L}(\bm{c}^{[t]})$\;
	}
	
	\caption{Structure of Hybrid Estimator.}
	\label{alg:hybrid}
\end{algorithm}

\subsection{Simulations}
\label{sec:doasim}
To assess the performance of the previously presented algorithms, we provide some simulation results. The considered system consists of $M=9$ omnidirectional antennas that form a uniform circular array (UCA). For simplicity, we assume that all of the $L=3$ sources lie in the same horizontal plane as the antenna array, such that the steering vector of the UCA only depends on the azimuth. In our simulations, we fix the ratio of the array radius $R$ and wavelength $\lambda$ to be equal to $1$. The switching network selects $K=4$ subarrays consisting of $W=3$ antennas according to the configuration given in Table~\ref{tab:subsampling}, which uses a clockwise numbering of the antenna elements of the UCA.

\begin{table}
	\caption{Subarray Sampling Scheme}\label{tab:subsampling}
	\begin{center}
		\begin{tabular}{c|c}
			 $k$ & Antenna Elements\\\hline
			 $1$ & $1,\;2,\;9$\\
			 $2$ & $1,\;3,\;8$\\
			 $3$ & $1,\;4,\;7$\\
			 $4$ & $1,\;5,\;6$
		\end{tabular}
	\end{center}
\end{table}

The parameters for the training of our algorithms, as well as the MCENet parameters, have been chosen according to Table \ref{tab:data}.\footnote{The MCENet parameters have been determined by a small scale random search. For details see Appendix~\ref{app:randomsearch}.} For the test set data, we use received signal realizations stemming from equally powered signals. Note that the parameters for the signal and noise variances in the training set have been chosen such that the resulting parameter space covers a reasonably broad operating range for the DoA estimation task. The knowledge about this limited parameter space is not incorporated in the model-based algorithms. However, the MCENet is trained with data from this range, which might introduce a certain advantage. By choosing the range broad enough, we want to make sure that this advantage is not too significant. 

\begin{table}
	\caption{Simulation Parameters DoA Estimation}\label{tab:data}
	\begin{center}
		\begin{tabular}{c|c}
			Parameter & Value\\\hline
			$\sigma_{s,\text{min}}^2$ & $-9\,\text{dB}$\\
			$\sigma_{\eta,\text{min}}^2$ & $-10\,\text{dB}$\\
			$\sigma_{\eta,\text{max}}^2$ & $30\,\text{dB}$\\
			$N_{\text{h}}$ & $4$\\
			$N_{\text{u}}$ & $2048$\\
			Weight Initialization & Glorot\cite{Glorot2010}\\
			Batch Size & $256$\\
			Optimizer & Adam\cite{Kingma2014}\\
			Learning Rate & $10^{-4}$\\
			Samples per Training Set & $128\times 10^6$\\
			Samples per Test Set & $10^3$
		\end{tabular}
	\end{center}
\end{table}

We denote the SSR method, for which we use YALMIP \cite{Lofberg2004} in combination with the MOSEK solver \cite{mosek} to solve the internal SOCP, simply by ``SSR'', whereas the alternating optimization variant (see Appendix \ref{sec:altopt}) is referred to by ``SSR iter.''. The ``SSR iter.'' variant uses a fixed number of update steps, which we set to $10^4$. For the presented SSR variants, we chose an oversampling factor of $32$, i.e., we use $32M$ equidistant grid points to cover the whole field of view. In contrast, the oversampling factor for the GLS method is set to $8$, because otherwise the computational complexity for $L=3$ sources becomes prohibitively large. As a reference, we added the results for a Genie ML approach. This approach starts with an initialization of the parameter vector $\bm{c}$ with the true DoAs, noise power, and covariance matrix of the transmit signals. From there, we use a block coordinate ascent to find the closest local maximum of the SML likelihood function, and use the DoAs corresponding to this maximum as the Genie ML estimates. Note that at high SNR the variance of the distance between the true DoAs and the local maximum of the likelihood function, i.e., the Genie ML estimates, is given by the Cram\'er-Rao-Bound (CRB), as the ML estimator is asymptotically efficient. However, the CRB is, as a stochastic bound on the error variance, not really applicable if only one noise and signal realization is paired with each DoA realization, unless an immense number of realizations is considered. Additionally, without enforcing any minimal distance between the DoAs, the calculation of the CRB suffers from numerical issues for closely spaced angles. Therefore, this Genie ML estimator gives a more reasonable performance bound than the CRB for our simulations. Furthermore, we show some results for the MVDR estimator \cite{Rieken2004}, which is---in contrast to the MUSIC approach presented in the same paper---technically applicable to scenarios with $L\geq W$ as well. However, note that the lack of a noise subspace in the subarray sample covariance matrices leads to the fact that $\hat{\bm{R}}_{\bm{y}}^{(k),-1}$ has no large eigenvalues that lead to a large $\bm{a}^\He\bm{G}^{(k),\He}(\theta)\hat{\bm{R}}_{\bm{y}}^{(k),-1}\bm{G}^{(k)}\bm{a}(\theta)$ for the directions $\theta$ without a source (in contrast to $L<W$, where $\bm{R}_{\bm{y}}^{(k),-1}$ converges to the projector onto the noise subspace [33]). Therefore, we expect that the MVDR spectrum cannot differentiate well between directions with and without sources in this scenario.

\subsubsection{Uncorrelated Sources}
\begin{figure}
	\begin{center}
		\includegraphics{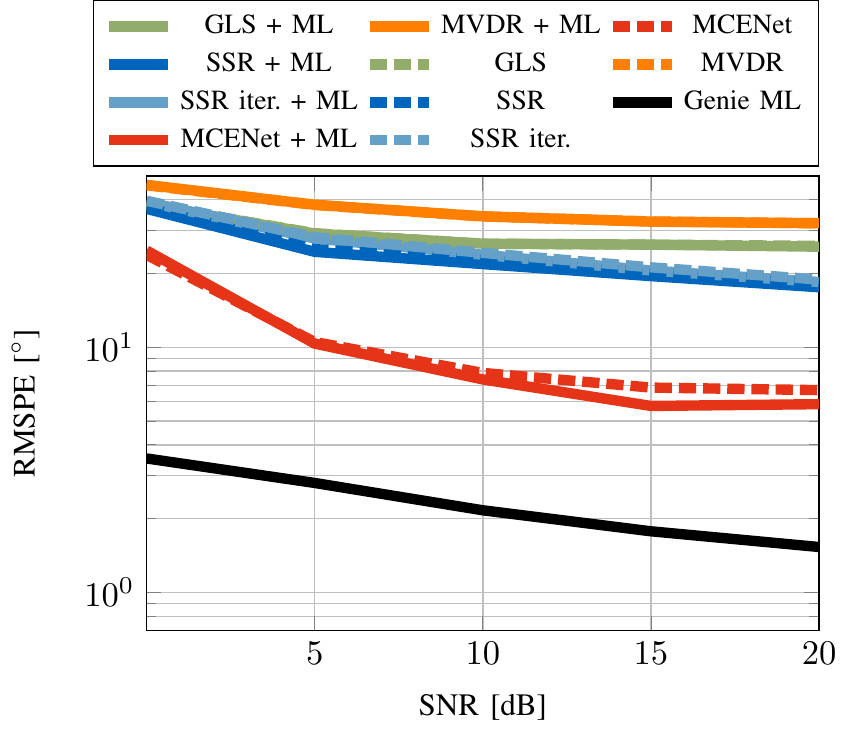}
	\end{center}
	\vspace{-15pt}
	
	\caption{RMSPE vs.~SNR, $N=10$.}
	\label{fig:rmse100}
\end{figure}

In Fig.~\ref{fig:rmse100}, we depict the root MSPE (RMSPE) of the different DoA estimators over the SNR, defined as $1/\sigma_\eta^2$, for $N=10$ snapshots. Thereby, the RMSPE under the assumption that the true and estimated DoAs are associated according to (\ref{eq:permutatedcost}) is given by
\begin{equation}
\text{RMSPE}=\sqrt{\frac{1}{L}\sum_{\ell=1}^L \Exp_\theta\left[\left|\modulo_{[-\pi,\pi)}\left(\theta_\ell-\hat{\theta}_\ell\right)\right|^2\right]}.
\end{equation}
The plot shows the hybrid approaches consisting of an initialization step and block coordinate ascent on the SML likelihood as solid lines and the plain results of the discussed methods without subsequent gradient steps as dashed lines. We can see that none of the proposed methods comes close to the performance of the Genie ML. Nevertheless, we can identify a large advantage of the MCENet approach over the model-based approaches.

\begin{figure}
	\begin{center}
		\includegraphics{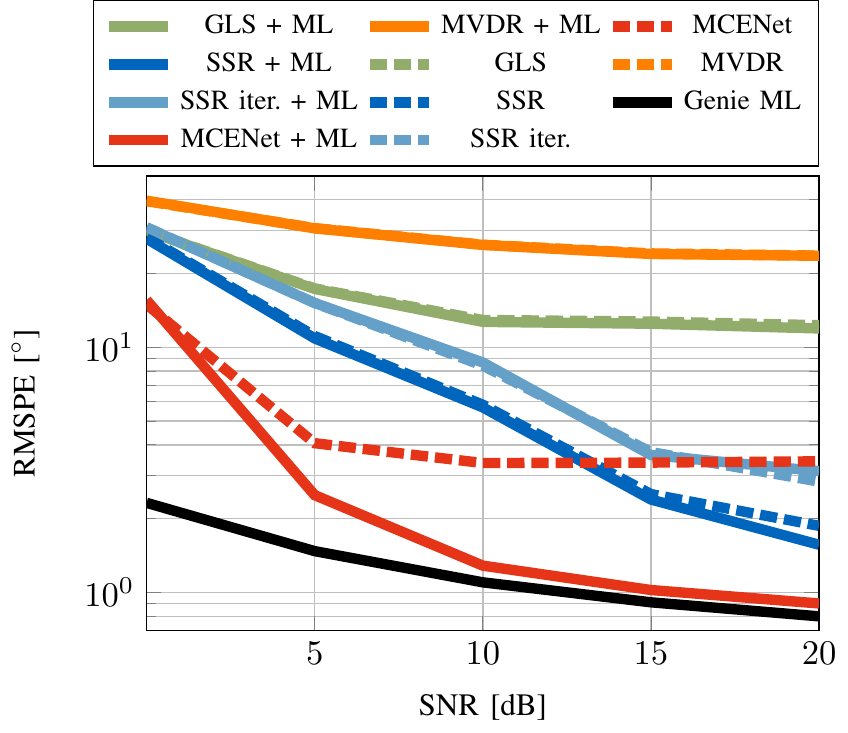}
	\end{center}
	\vspace{-15pt}
	
	\caption{RMSPE vs.~SNR, Top $90\%$ of Realizations, $N=10$.}
	\label{fig:rmse90}
\end{figure}

\begin{figure}
	\begin{center}
		\includegraphics{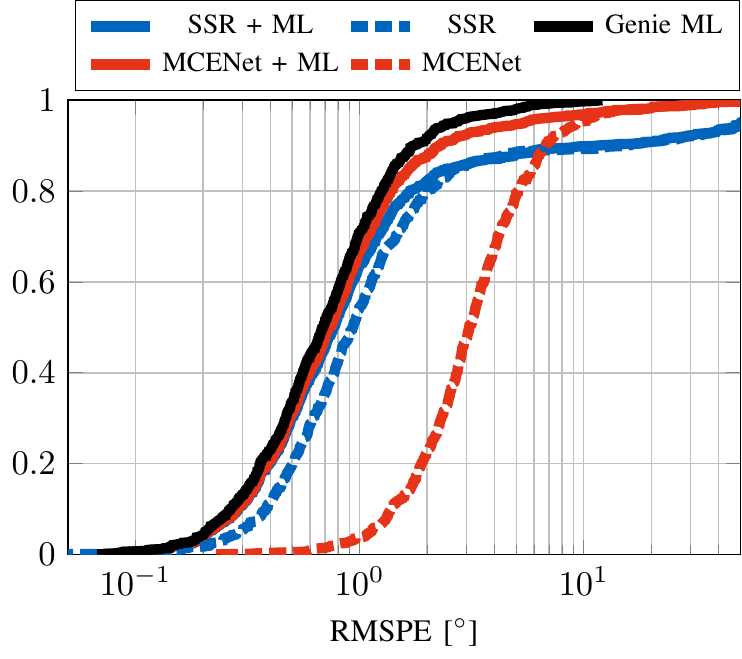}
	\end{center}
	\vspace{-15pt}
	
	\caption{Empirical Cumulative Density Function, SNR$=20\,\text{dB}$, $N=10$, $M=9$.}
	\label{fig:cdfN10}
\end{figure}

To understand where this performance advantage of the MCENet stems from, we plot the RMSPE of the top $90\%$ of realizations for each DoA estimator in Fig.~\ref{fig:rmse90}. Now we can see that the hybrid MCENet approach almost achieves the performance of the Genie ML for an SNR of $10\,\text{dB}$ and higher. Meanwhile, the model-based approaches are still falling behind. Interestingly, $10^4$ update steps are not enough for the alternating optimization approach ``SSR iter.'' to achieve the same performance as the general purpose solver solution. In last place are the GLS approach that might perform better for a denser grid, which, however, is computationally intractable, and the MVDR technique. This reduced gap between the GenieML performance and the performance of the other algorithms shows that the results in Fig.~\ref{fig:rmse100} are dominated by the suboptimal performance for some of the realizations, which we will refer to as outliers, whereas for the majority of the realizations the algorithms achieve an acceptable accuracy. The hybrid MCENet approach suffers from fewer outliers than the model-based approaches, which can be seen in Fig.~\ref{fig:cdfN10} as well, where we plotted the empirical cumulative density of the RMSPE per DoA estimator at $20\,\text{dB}$ SNR for the Genie ML, the SSR and the MCENet methods.

Interestingly, the presented estimators differ in which realizations lead to outliers, e.g., a realization that leads to an outlier for the MCENet NN does not necessarily result in an outlier for the SSR method or vice versa. Whereas our simulations showed that the GLS estimator struggles especially with closely spaced sources, which is related to the very coarse grid that it is based on, categorizing which kind of realizations lead to outliers for the MCENet and SSR methods turned out to be unsuccessful. For the MCENet NN, the realizations that lead to outliers depend on the training data, which makes a theoretical analysis of these critical case rather difficult and shall be left to later investigations. Specific opportunities to compensate for these deficits may arise through more sophisticated active learning strategies. Similarly, the SSR method is based on the solution of an optimization problem, which makes it difficult to obtain an intuition about the nature of the outlier realizations.

When we compare the hybrid MCENet results with the plain MCENet output, we see in both Fig.~\ref{fig:rmse90} and Fig.~\ref{fig:cdfN10} that the combination of the data-based MCENet with the model-based gradient steps is crucial. The MCENet alone cannot provide estimates that can compete with a model-based approach for non-outlier realizations. This comes as a trade-off between high SNR accuracy and outlier robustness, which is also reflected in the cost function of the NN. By design, the NN tries to minimize the average MCE over all realizations. In that sense, minimizing the occurrence of outliers that lead to large errors weighs more than further improving the accuracy for realizations with a small error such that during training the emphasis lies first and foremost on the robustness against outliers.

\begin{figure}
	\begin{center}
		\includegraphics{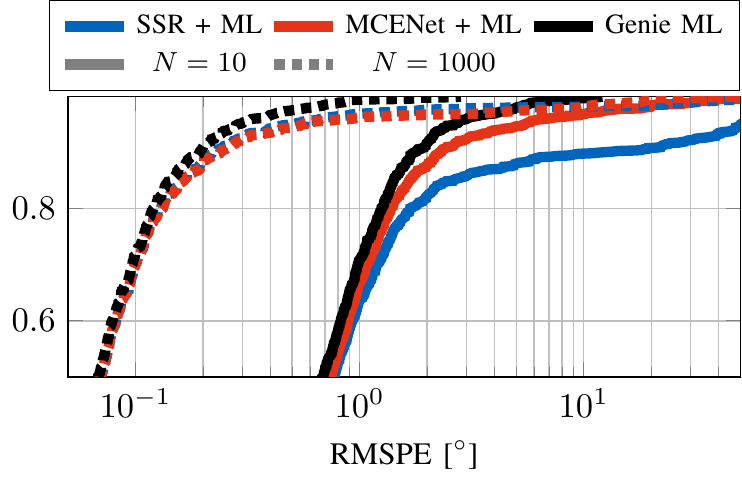}
	\end{center}
	\vspace{-15pt}
	
	\caption{Empirical Cumulative Density Function, SNR$=20\,\text{dB}$, varying $N$, $M=9$.}
	\label{fig:cdfN10vs1000}
\end{figure}

The higher susceptibility to outliers of the model-based approaches vanishes for a higher number of snapshots $N$, as can be seen in Fig.~\ref{fig:cdfN10vs1000}. There, we compare the relevant cut-out of the empirical cumulative density of the hybrid approaches for $N=10$ and $N=1000$. For high $N$, we see not only a general shift of the CDF curves (the CRB scales linearly in $N$), but almost no outliers occur and the performance of SSR is on par with the MCENet. This result is not surprising, as the SSR method is based on a covariance-matching criterion. The sample covariance matrices are consistent estimates of the true subarray covariance matrices, which in turn justifies the validity of the covariance-matching objective for high $N$. Note that the GLS estimator, which is based on a similar objective as the SSR method, has been proven to be a consistent estimator (for a sufficiently dense grid) \cite{Sheinvald1999}.

\begin{figure}
	\begin{center}
		\includegraphics{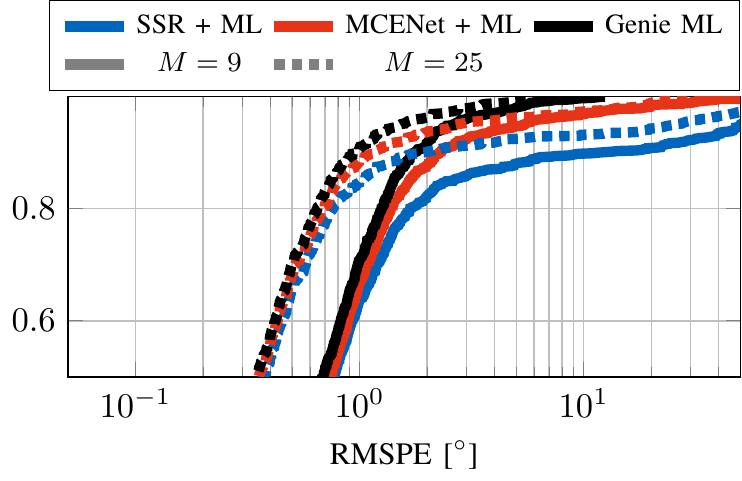}
	\end{center}
	\vspace{-15pt}
	
	\caption{Empirical Cumulative Density Function, SNR$=20\,\text{dB}$, $N=10$, varying $M$.}
	\label{fig:cdfM10}
\end{figure}

An increase in the number of antennas $M$ improves the average performance of all discussed methods, as is shown in Fig.~\ref{fig:cdfM10}. However, for $M=25$ antennas, we still see a significant amount of outliers for the MCENet and SSR approaches. Furthermore, increasing $M$ not only comes with additional hardware expenditures, but the number of subarrays $K$, which have to be sampled, and therefore, the time to scan the whole array, grows as well.

For an assessment of the case of strictly more sources than RF chains, the considered UCA with $W=3$ RF chains is not well suited, because the estimation task for $L=4$ sources and the considered system becomes so difficult that the MCENet and SSR methods are not able to produce reasonable estimates for a majority of realizations. Instead, we were able to find encouraging results for other array geometries, but due to the lack of space further details are beyond the scope of this work.

Another important result to be stated is when referring to the generalization capabilities of the proposed data-based method, i.e., how well the MCENet NN generalizes to data outside the parameter ranges which the training data has been drawn from. Here it is important to understand that data-based approaches in general can only perform well in cases that have been sufficiently represented during the training phase. A similar observation can be made in the application at hand. In cases where the proposed method is applied to data which is outside the specified parameter ranges, the proposed estimator suffers from a severe degradation. However, a proper experimental design of the training phase can easily prevent from those cases. Which is obviously a disadvantage in comparison to model-based approaches turns out to be advantageous in cases where an appropriate stochastic model cannot be easily derived and model-based approaches suffer from an inevitable model mismatch.

\subsubsection{Correlated Sources}
\begin{figure}
	\begin{center}
		\includegraphics{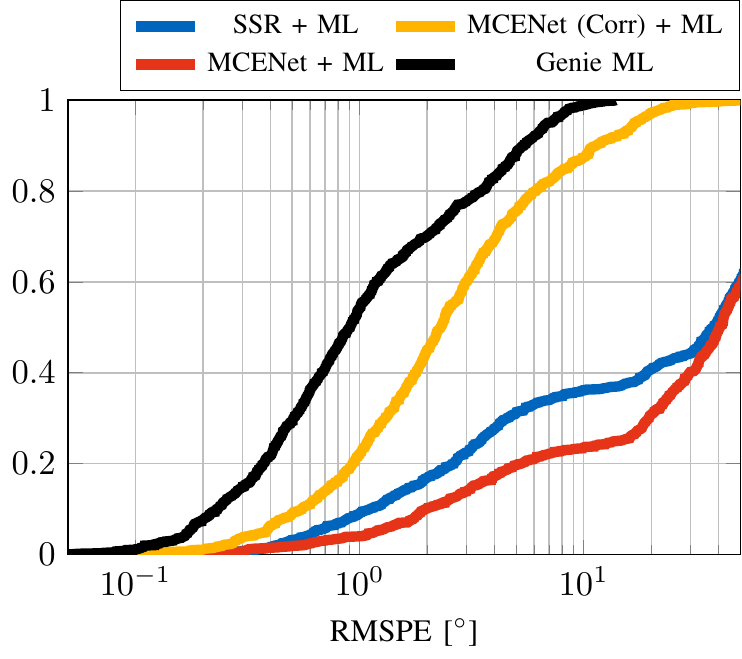}
	\end{center}
	\vspace{-15pt}
	
	\caption{Empirical Cumulative Density Function for Correlated Sources, SNR$=20\,\text{dB}$, $N=10$, $M=9$, $\rho=1$.}
	\label{fig:cdfCorr}
\end{figure}

\begin{figure}
	\begin{center}
		\includegraphics{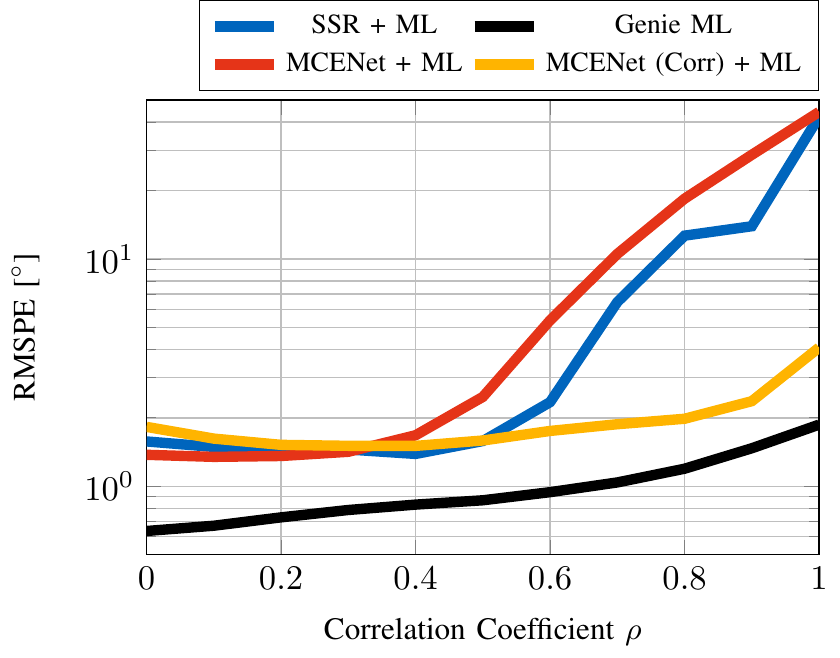}
	\end{center}
	\vspace{-15pt}
	
	\caption{RMSPE vs.~Correlation Coefficient $\rho$, Top $90\%$ of Realizations, $N=10$.}
	\label{fig:corr}
\end{figure}

A prime example for the aforementioned adaption of the NN-based estimator by augmenting the training data set arises when we take a look at correlated sources. To that end, we created correlated realizations with a transmit covariance matrix 
\begin{equation}
\bm{R}_{\bm{s}} = \begin{bmatrix}
1 & \rho & \rho^2\\
\rho & 1 & \rho\\
\rho^2 & \rho & 1,
\end{bmatrix}
\end{equation}
with $\rho\in[0,1]$, to compare the performance of the presented estimators. The results at an SNR of $20$\,dB can be found in Fig.~\ref{fig:cdfCorr} for $\rho = 1$ and in Fig.~\ref{fig:corr} for varying $\rho$. In addition to the MCENet NN that has been trained on data with uncorrelated sources, we also added the results for an MCENet NN that has been trained with correlated sources denoted by ``MCENet (Corr)''. For the training of the MCNet (Corr), we not only sampled the transmit power differences between the sources from a uniform distribution between $0$\,dB and $-9$\,dB, but also the correlation coefficient $\rho$ followed a uniform distribution between $0$ and $1$. This data generating model for the training set of the MCENet (Corr) comprises the realizations of the test set with correlated sources. Furthermore, we use the MCENet NN trained on uncorrelated sources as a starting point for our training procedure of the MCENet (Corr) NN. From this initialization, we used 16 million data samples stemming from the correlated data model to adapt the existing network to the new scenario. This adaptation method resembles an online training procedure as has been discussed in the context of model order selection in \cite{Barthelme2020}. In contrast, the SSR method is inherently based on the uncorrelated data model and cannot be easily adapted to the correlated case.

We see in Fig.~\ref{fig:cdfCorr} that in the fully correlated case, the SSR and MCENet method, which has been trained on uncorrelated data, fail to provide reasonable estimates. This is due to the fact that the SSR method is based on the uncorrelated source model and the MCENet NN operates on data that is very different from its training data. The MCENet (Corr) NN, however, shows a much better performance, although it is also not able to fully achieve the Genie ML bound. The increased generalization
capabilities come at the price of slightly degraded performance for correlation coefficients close to zero (cf.\ Fig.~\ref{fig:corr}). The slight performance loss of MCENet (Corr) at low $\rho$ is the result of a trade-off between a high accuracy at low and high correlation. This reflects the MCENet training objective, which penalizes large estimation errors in the training set, and therefore, promotes robustness towards critical cases at the cost of a reduced accuracy in easier scenarios. 

\subsubsection{Fully Sampled Array} 
\begin{figure}
	\begin{center}
		\includegraphics{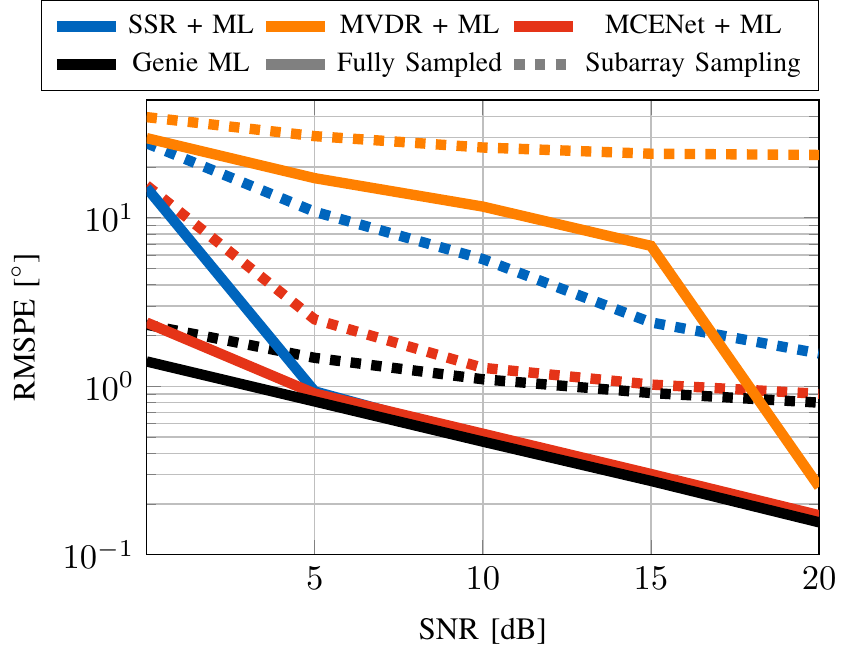}
	\end{center}
	\vspace{-15pt}
	
	\caption{RMSPE vs.~SNR for the Fully Sampled Array, Top $90\%$ of Realizations, $N=10$.}
	\label{fig:rmse90full}
\end{figure}

The MCENet NN can be trained not only for arrays with subarray sampling, but also for fully sampled arrays. In fact, the fully sampled scenario represents a special case of the systems with subarray sampling for $K=1$ and $W=M$. Again, the estimator outputs are not completely free of outliers. Hence, we again show the RMSPE of the top $90\%$ of realizations for each estimator in Fig.~\ref{fig:rmse90full}. As a comparison, we added the results for the case with subarray sampling as dashed lines. Note that although we could choose the same realizations for the DoAs, the signal and noise realizations for both cases differ from each other because of the different dimensionality of both cases.

Fig.~\ref{fig:rmse90full} shows that for a fully sampled array, the performance gap between the MCENet approach and the SSR reference almost vanishes. The performance of both methods only differs for very low SNR, where the MCENet provides a little more robustness. Also for high SNR, the number of outliers experienced by the MVDR method decrease significantly compared to the more complex case of subarray sampling. However, note that the estimators do not only differ in their achievable accuracy, but also in their computational complexity.

\subsubsection{Complexity}
On a final note, we want to briefly discuss the complexity of the presented estimators. To this end, we show in Table~\ref{tab:doatime} the computation times of the individual estimators with and without consecutive gradient steps for uncorrelated sources and the UCA with subarray sampling. The presented times are for $1000$ realizations at $20\,\text{dB}$ SNR in {MATLAB} on a simulation server equipped with two Intel Xeon Gold 6134 processors. Although we know that computation times do not achieve the same validity as a rigorous complexity analysis in Landau notation, due to their dependence on the used hardware and implementation, they still yield some qualitative insights. From our simulations, we see that the MCENet inference steps took about a tenth of the evaluation time of the ``SSR'' estimator, which itself is again about ten times faster than the ``SSR iter.'' with a fixed iteration count of $10^4$ iterations and the GLS estimator, whose complexity grows exponentially with the number of sources. Only the MVDR estimator, which has the worst accuracy, is able to compete with the MCENet approach in terms of computational complexity. Taking the consecutive gradient steps into account\footnote{Note that the required time for the gradient steps heavily depends on the target accuracy. A looser stopping criterion may significantly reduce the required computation times. For the presented simulations, the stopping criterion for the gradient steps is very tight ($<10^{-6}$ absolute change in the log-likelihood), to achieve meaningful results for the MDL approach discussed in the next section.}, we see that these steps implemented by a block coordinate ascent take roughly the same time for all discussed methods. 
\begin{table}
	\caption{Computation Times of DoA Estimators}\label{tab:doatime}
	\begin{center}
		\begin{tabular}{c|c c c c c c}
			& MCENet & SSR & SSR iter. & GLS & MVDR\\\hline
			non-hybrid & $5.4\,\text{s}$ & $127.2\,\text{s}$ & $1002.8\,\text{s}$ & $1486.2\,\text{s}$ & $6.5\,\text{s}$\\
			hybrid & $52.3\,\text{s}$ & $174.7\,\text{s}$ & $1045.3\,\text{s}$ & $1541.8\,\text{s}$ & $53.8\,\text{s}$
		\end{tabular}
	\end{center}
\end{table}

\section{Model Order Selection}
\label{sec:modelorder}
Knowledge about the number of sources in the transmission environment $L$ is essential in any of the previously presented DoA estimation approaches. With an inaccurate estimate of the model order, we base our algorithms on the wrong stochastic model or choose the wrong NN, which has been trained on mismatched data. Hence, we discuss the problem of model order selection in this section. Again, we follow the structure of the previous section and discuss model-based approaches first, namely information criteria (IC). Then we present a data-based approach, which uses a classification NN, and a performance comparison based on Monte Carlo simulations.
\subsection{Information Criteria}
The most common model-based model order selection methods are based on ICs \cite{Stoica2004}. All variants of these IC follow a common structure of their underlying optimization problem. For the considered system model this optimization problem is given by
\begin{equation}
	\hat{L}=\argmax_{\ell\in\{0,\dots,L_{\text{max}}\}} \ln\left(p_\ell\left(\bm{Y};\hat{\bm{\theta}},\hat{\bm{R}}_{\bm{s}},\hat{\sigma}_\eta^2\right)\right) + c(\ell),
	\label{eq:infocriteria}
\end{equation}
where $\bm{Y}$ contains all observations $\bm{y}^{k}(n),k=1,\dots,K,n=1,\dots,N$, the likelihood function of the received signals under the assumption that the model order is equal to $L$ and parameterized by the ML estimates of the model parameters is denoted by $p_\ell(\cdot)$, and $c(\ell)$ is a penalty term that combats overfitting of the model order.

In the fully sampled case, the likelihood function can be reparameterized by the eigenvalues of the sample covariance matrix. This leads to a very convenient expression for the value of the likelihood function that depends only on these eigenvalues and no longer on the DoA estimates for each considered model order \cite{Wax1985}. Therefore, the computational load is basically reduced to an eigenvalue decomposition in contrast to evaluating ML estimates for very high model orders up to $L_{\text{max}}$. Unfortunately, this reparameterization is no longer available when we consider systems with subsampling. This is made visible by looking at eigendecompositions of the sample covariances $\hat{\bm{R}}_{\bm{y}}^{(k)}$, where the true model order $L$ is larger than the number of RF chains $W$. Here, the eigenspace can no longer be decomposed into a signal and noise subspace. Additionally, as we see from the discussion in Section \ref{sec:doa}, the ML estimates that are generally needed for the evaluation of the IC cannot be obtained directly for $L\geq W$.

Instead, we can replace the ML estimates of the model parameters in (\ref{eq:infocriteria}) by the GLS estimates, as has been proposed in \cite{Haykin1995}. Applying the same rational, the SSR estimator or any hybrid version can be used to evaluate the IC.

\subsection{Purely Data-Based Model Order Selection}
As we have seen in Section \ref{sec:doasim}, the DoA estimates in the low SNR and low number of snapshots region are heavily affected by outliers. In \cite{Barthelme2020}, it is shown that in the fully sampled case a NN-based model order selection approach can outperform classical IC in exactly this region, while simultaneously performing equally for high SNR and many snapshots. Therefore, we follow the lines of \cite{Barthelme2020} and discuss a similar NN, to which we refer to as \emph{CovNet}, for model order selection for systems with subsampling. 
\subsubsection{Data and preprocessing}
For the NN, we use the same kind of preprocessing based on the sample covariance matrices as described in Section \ref{sec:data}. Again, due to the artificial data we use, we can sample from the underlying stochastic model as described for MCENet. However, the network is now not only fed with data stemming from a stochastic model with fixed model order $L$, but we have to provide data for varying model orders $L=0,\dots,L_{\text{max}}$. This model order is used in the form of a one-hot encoded vector as the label for each data sample. During training, each batch consists of an equal number of realizations from the varying model orders such that no bias towards one model order is introduced. 

\subsubsection{Architecture and cost function}
Again, we use a fully connected, feedforward NN with $N_{\text{h}}$ hidden layers with $N_{\text{u}}$ neurons each and ReLU activation. The output layer consists of $L_{\text{max}}+1$ neurons and applies a softmax operation to form the outputs $z(\ell),\ell=0,\dots,L_\text{max}$ \cite{Bridle1989}. By training based on the cross-entropy loss, which for one-hot encoded labels is given by
\begin{equation}
\max_{\bm{w}}\ln\left(z(\ell^*|\bm{x};\bm{w})\right),
\label{eq:crossentropy}
\end{equation}
the output values $z(\ell)$ can be interpreted as estimates of the posterior probabilities for model order $\ell$. The training procedure can be seen as a heuristic approach to the optimal maximum a posteriori (MAP) estimator, because the training adapts the weights $\bm{w}$ such that the estimate of the posterior probability $z(L)$ of the true model order $L$ of the input $\bm{x}$ is maximized \cite{Lecun1998}. 

\subsection{Simulations}
\label{sec:modelordersim}
We conducted simulations for model order selection with the same data generating model as introduced in Section \ref{sec:doasim}. The maximal number of sources $L_{\text{max}}$ that we consider for our simulations is $3$, i.e., we are operating in the range of $L_{\text{max}}\geq W$. For CovNet, we use a smaller network than MCENet. CovNet has the same structure as its counterpart in \cite{Barthelme2020} with $N_{\text{h}}=3$ layers with $N_{\text{u}}=1024$ neurons and has been trained on $10^6$ batches with $64$ samples in each batch, by using the Adam optimizer \cite{Kingma2014} with fixed learning rate of $10^{-2}$. As a reference, we use the maximum description length (MDL) estimator, whose penalty term in (\ref{eq:infocriteria}) under the assumption of uncorrelated transmit signals is given as \cite{Haykin1995}
\begin{equation}
	c(\ell)=\frac{2\ell+1}{2}\ln\left(KN\right),
\end{equation}
and uses the hybrid SSR method, which is computationally tractable and achieves a better DoA estimation performance than the GLS approach (cf.~Section~\ref{sec:doasim}), to obtain the necessary parameter estimates.

\begin{figure}
\includegraphics{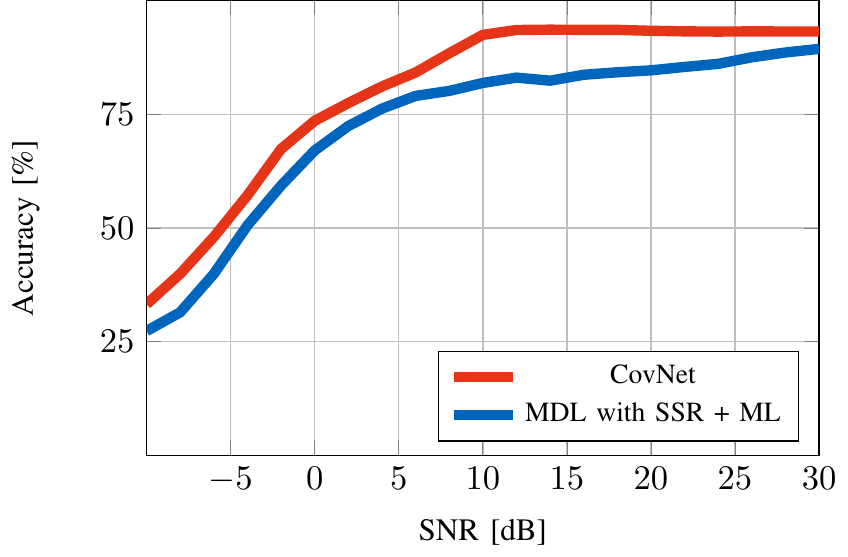}
\caption{Model Order Selection Accuracy vs.~SNR, $N=10$}
\label{fig:accuracySNR}
\end{figure}

\begin{figure}
\includegraphics{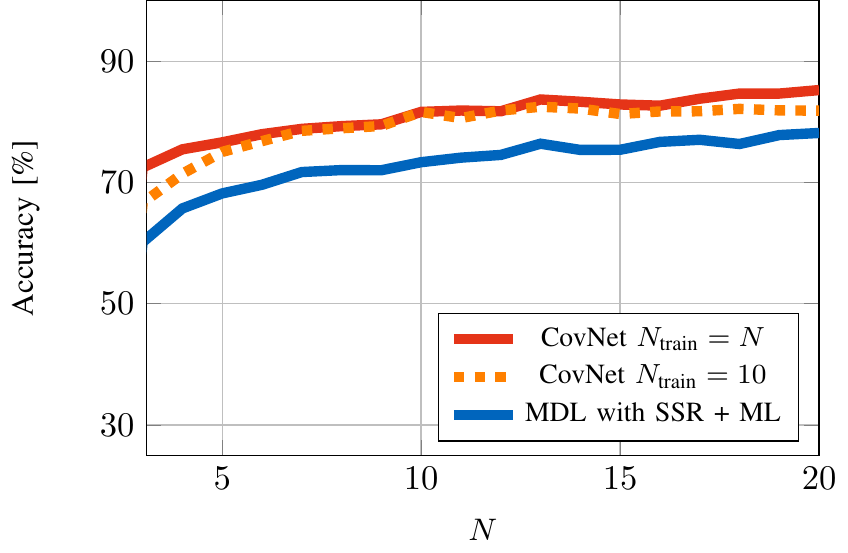}
\caption{Model Order Selection Accuracy vs. $N$}
\label{fig:accuracyN}
\end{figure}

In Fig.~\ref{fig:accuracySNR}, we show the model order selection accuracy of the discussed methods for varying SNR. To that end, we use a test set consisting of $4\cdot 10^3$ data samples with a fixed SNR and an equal number of data points from all model orders. Note that due to our SNR definition, fixed SNR means that the power ratio of the strongest source to the noise is constant, but the transmit powers of the weaker sources are still randomly drawn, i.e., the ratio of transmit power and noise power for these sources is smaller than the stated SNR. Similarly, we show the achieved accuracy of the different methods for a different number of snapshots $N$ in Fig.~\ref{fig:accuracyN}, where the respective test sets consist of data samples with a fixed number of snapshots $N$ and randomly drawn SNR. In both cases, CovNet achieves a significantly higher accuracy than the MDL estimator. As we are operating in a low snapshot region, the SSR estimators are prone to outliers, as discussed for $L=3$ in Section \ref{sec:doasim}, which leads to the suboptimal performance of the MDL estimator compared to the NN-based approach. 

Note that, in Fig.~\ref{fig:accuracyN}, we show two different CovNet results. The solid red line shows the accuracy for a NN, where the number of snapshots in the training set and test set are matching, i.e., $N_{\text{train}}=N$, whereas the dashed orange line shows the performance of a CovNet model that has been trained on data with $N_{\text{train}}=10$ snapshots, which means that for this model the data in the test and training sets are different. Interestingly, the CovNet model trained on $10$ snapshots is able to generalize well to data with a different number of snapshots and achieves almost the same performance over all $N$ as the NNs that have been trained on the same number of snapshots as in the test set $N=N_{\text{train}}$. This means that for the implementation in a direction finder, NNs for each possible number of snapshots do not have to be stored, but a certain realization can cover different $N$.

Again, we end this section by a short comparison of the required computation times of each presented model order selection algorithm. For $1000$ realizations with varying SNR evaluated on the same simulation server as discussed in Section~\ref{sec:doasim}, the inference from the CovNet model takes $2.6$ seconds. However, the MDL estimator takes $505.2$ seconds for the same task, since a DoA estimate for all possible model orders---including the computationally expensive high model orders---has to be performed for every realization. This is a difference by a factor of $200$. Although the CovNet approach does not automatically yield a DoA estimate like the MDL approach, its execution time combined with the time for a consecutive DoA estimation for the estimated model order (cf.~Section~\ref{sec:doasim} for $L=3$) is still significantly smaller.

\section{Conclusion}
From the simulation results in Section \ref{sec:doasim} and \ref{sec:modelordersim}, we see that NN-based approaches to DoA estimation and model order selection are viable alternatives to existing model-based techniques for systems with subarray sampling. In terms of selection accuracy and DoA estimation error, the proposed NN schemes are able to outperform model-based techniques when the number of available snapshots is small. Hereby, a combination of NN based initialization and model-based gradient steps was crucial to achieve competitive DoA estimates, although improvements on the architecture or training procedure may further improve the purely NN-based estimates (cf.~results in \cite{Bialer2019}). Additionally, the computational complexity of a NN inference is considerably lower than the evaluation of model-based estimators, which enables completely NN-based DoA estimation chains for time-critical applications. 

However, there are still some open questions that need to be addressed: How do NN based approaches cope with array calibration? And, how robust are these methods when model imperfections come into play? One idea to tackle these problems is to use an online learning procedure to adapt a pretrained NN to the changed model as has already been proposed in \cite{Barthelme2020}.

\begin{appendices}
\section{Alternating Update for SSR Estimator}
\label{sec:altopt}
The optimization problem for the SSR estimator, according to \cite[Equation (46)]{Suleiman2018}, is given by
\begin{equation}
	\begin{split}
	\min_{\bm{p},\sigma_\eta^2}\,&\sum\limits_{k=1}^K\trace\left(\check{\bm{R}}_{\bm{y}}^{(k),-1}\hat{\bm{R}}_{\bm{y}}^{(k)}\right)\\
	\st &\,\bm{p}\geq\mathbf{0},\sigma_\eta^2\geq 0,\\
	&\,\sum\limits_{g=1}^{G}w_g p_g+\bar{w}\sigma_\eta^2=1,
	\end{split}\label{eq:ssropt}
\end{equation}
with the sparse representation of the covariance matrix
\begin{equation}
\check{\bm{R}}_{\bm{y}}^{(k)}=\check{\bm{A}}^{(k)}\diag\{\mathbf{p}\}\check{\bm{A}}^{(k),\He}+\sigma_\eta^2\mathbf{I},
\end{equation}
where $\check{\bm{A}}^{(k)}$ is a dictionary containing $G$ subarray steering vectors $\bm{G}^{(k)}\bm{A}(\check{\bm{\theta}}_g)$, $g=1,\dots,G$, and $\bm{p}$ contains the respective power values.

The weights in (\ref{eq:ssropt}) are given as 
\begin{align}
	w_g&=\frac{1}{KW}\sum\limits_{k=1}^K\bm{a}^\He(\check{\bm{\theta}}_g)\bm{G}^{(k),\He}\hat{\bm{R}}_{\bm{y}}^{(k),-1}\bm{G}^{(k)}\bm{a}(\check{\bm{\theta}}_g),\\
	\bar{w}&=\frac{1}{KW}\sum\limits_{k=1}^K\trace\left(\hat{\bm{R}}_{\bm{y}}^{(k),-1}\right).
\end{align}
Note that we added a missing factor of $1/K$ compared to \cite[Equation (46)]{Suleiman2018}, as (cf.~\cite[Equation (17)]{Stoica2011})
\begin{equation}
\begin{split}
	\sum_{k=1}^{K}\sum_{g=1}^{G}p_g\bm{a}^\He(\check{\bm{\theta}}_g)\bm{G}^{(k),\He}\hat{\bm{R}}_{\bm{y}}^{(k),-1}\bm{G}^{(k)}\bm{a}(\check{\bm{\theta}}_g)\\+\sum_{k=1}^{K}\sigma_\eta^2\trace\left(\hat{\bm{R}}_{\bm{y}}^{(k),-1}\right)\xrightarrow[N \to \infty]{}KW.
	\end{split}
\end{equation}

Following the lines of \cite[Section III]{Stoica2011}, we obtain the alternating update rules in the $i+1$-th iteration as
\begin{align}
	p_g^{[i+1]}&=p_g^{[i]}\frac{\left\|\sum\limits_{k=1}^{K}\bm{a}^\He(\check{\bm{\theta}}_g)\bm{G}^{(k),\He}\check{\bm{R}}_{\bm{y}}^{(k),-1}\hat{\bm{R}}_{\bm{y}}^{(k),1/2}\right\|_2}{w_g^{1/2}\xi^{[i]}},\\
\sigma_\eta^{2,[i+1]}&=\sigma_\eta^{2,[i]}\frac{\left(\sum\limits_{k=1}^{K}\trace\left(\check{\bm{R}}_{\bm{y}}^{(k),-1}\hat{\bm{R}}_{\bm{y}}^{(k)}\check{\bm{R}}_{\bm{y}}^{(k),-1}\right)\right)^{1/2}}{\bar{w}^{1/2}\xi^{[i]}},
\end{align}
with
\begin{equation}
\begin{split}
	\xi^{[i]}=&\sum\limits_{g=1}^Gw_g^{1/2}p_g^{[i]}\left\|\sum\limits_{k=1}^{K}\bm{a}^\He(\check{\bm{\theta}}_g)\bm{G}^{(k),\He}\check{\bm{R}}_{\bm{y}}^{(k),-1}\hat{\bm{R}}_{\bm{y}}^{(k),1/2}\right\|_2\\
	&+\bar{w}\sigma_\eta^{2,[i]}\sum\limits_{k=1}^{K}\trace\left(\check{\bm{R}}_{\bm{y}}^{(k),-1}\hat{\bm{R}}_{\bm{y}}^{(k)}\check{\bm{R}}_{\bm{y}}^{(k),-1}\right).
\end{split}
\end{equation}

\section{Justification of MCENet Parameters}
\label{app:randomsearch}
Let us start by stating that the architecture and parameters of the MCENet NN have not been heavily optimized with regard to its performance shown in Section \ref{sec:doasim}. Clearly, a more sophisticated architecture (e.g., with convolutional layers) or an exhaustive search in the hyperparameters has the potential to improve the performance, as is the case for any NN-based approach. Nevertheless, we are confident that the hyperparameters, which we chose for the MCENet NN, lead to a fair assessment of the achievable performance of the presented estimator. The reason behind our evaluation lies in the results of a small scale random search \cite{Bergstra2012}, which we performed to select the hyperparameters of the network. To that end, we trained multiple MCENet NNs with randomly chosen hyperparameters from a certain range with data from the same training set and compared their achieved loss on a common cross-validation data set.\footnote{Since we effectively train the networks for a single epoch on a very large training set, we do not need additional measures to combat overfitting like early stopping.} In particular, we sampled the number of hidden layers $N_{\text{h}}\in\{1,2,3,4\}$, the number of neurons per hidden layer $N_{\text{u}}\in\{128,256,\dots,1920,2048\}$, and the initial learning rate from the set $\{10^{-2},10^{-3},10^{-4}\}$. Hereby, an initial learning rate of $10^{-4}$ showed the best convergence of the Adam optimizer out of the three choices.

\begin{figure}
	\begin{center}
		\includegraphics{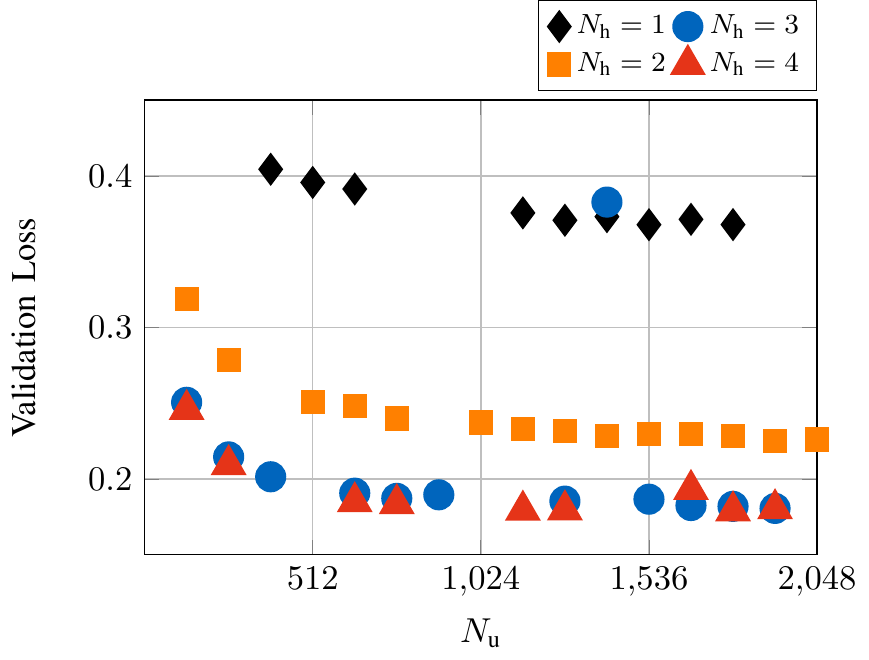}
	\end{center}
	\vspace{-15pt}
	
	\caption{Random Search Results - Validation Loss vs.\ Number of Neurons per Layer.}
	\label{fig:TrainingValidation1}
\end{figure}

The results of the random search for a learning rate of $10^{-4}$ can be found in Fig.~\ref{fig:TrainingValidation1}. There, we show how the achieved validation loss depends on the number of neurons $N_{\text{u}}$. Note that some points in Fig.~\ref{fig:TrainingValidation1} are missing as the employed random search does train a NN only for a random subset of combinations of hyperparameters instead of each possible combination. For each choice of $N_{\text{h}}$, the validation loss generally decreases for growing $N_{\text{u}}$, especially for a low number of neurons. However, the rate at which the validation loss improves is getting smaller as well such that we see a saturation of the performance in the number of neurons. Moreover, we can see that the validation loss does not significantly improve between choosing $3$ or $4$ hidden layers. Therefore, we come to the conclusion that a further increase in the number of hidden layers or neurons does not yield a significant performance gain.

\end{appendices}
\bibliographystyle{IEEEtran}
\bibliography{DoALitDB}

\end{document}